\documentclass[twocolumn,showpacs,preprintnumbers,amsmath,amssymb]{revtex4}

\usepackage[unicode]{hyperref}
\usepackage[caption=false]{subfig}
\usepackage{graphics}
\usepackage{epstopdf}
\usepackage{epsfig}
\usepackage{graphicx}
\graphicspath{ {./images/} }
\usepackage{dcolumn}
\usepackage{bm}
\usepackage[caption=false]{subfig}

\begin{document}

\title{Product-Sum universality and Rushbrooke inequality in explosive percolation 
}%

\author{ M. K. Hassan and M. M. H. Sabbir 
}%
\date{\today}%

\affiliation{
University of Dhaka, Department of Physics, Theoretical Physics Group, Dhaka 1000, Bangladesh \\
}

\begin{abstract}
We study explosive percolation (EP) on Erd\"{o}s-R\'{e}nyi network for product rule (PR) and sum rule 
(SR). Initially, it was claimed that EP describes discontinuous phase transition, now it is
well-accepted as a probabilistic model for thermal continuous phase transition (CPT). However, no model 
for CPT is complete unless we know how to relate its observable quantities with those of thermal CPT.
To this end, we define entropy, specific heat, re-define
susceptibility and show that they behave exactly like their thermal counterparts.
 We obtain the critical exponents $\nu, \alpha, \beta$ and $\gamma$ numerically
and find that both PR and SR belong to the same universality class and they obey 
Rushbrooke inequality.

\end{abstract}


\maketitle

The notion of percolation is omnipresent in many seemingly disparate natural and man-made systems 
\cite{ref.Stauffer}. Examples include spread of forest fire, flow of 
fluid through porous media, spread of biological and computer viruses etc.
 \cite{ref.saberi, ref.Newman_virus, ref.Moore_virus}. 
Besides such direct applications, percolation
is best known as a paradigmatic model for phase transition.
One of the simplest models for percolation is the classical random percolation (RP) on 
Erd\"{o}s-R\'{e}nyi (ER) network
in which one starts with $N$ labeled nodes that are initially all isolated \cite{ref.erdos}. 
Then at each step a link, say $e_{ij}$, is picked at random from all the possible pair of links   
and occupy it to connect nodes $i$ and $j$. 
As the number of occupied links $n=tN$ increases from zero we find that clusters, i.e. contiguous 
nodes connected by occupied links, are formed and on the average grown.
In the process, the largest cluster $s_{{\rm max}}$ undergoes a transition across $t_c=0.5$ from 
minuscule size ($s_{{\rm max}}\sim \log N$)  to  giant size ($s_{{\rm max}} \sim N$).
The emergence of such threshold value $t_c$ is found to be accompanied by a sudden change in the order parameter $P$, 
the ratio of the largest cluster to the network size, such that $P=0$ at $t\leq t_c$ and $P>0$ at $t>t_c$ in the limit $N\rightarrow \infty$. This is 
reminiscent of the second order or continuous phase transition (CPT).

In 2009, Achlioptas {\it et al.} proposed a class of percolation model in which two links  
are picked randomly instead of one at each step \cite{ref.Achlioptas} . However, ultimately only one of the links, 
that results in the smaller clustering, is occupied and the 
other is discarded for future picking. One of the key features of this rule, which is now known as the 
Achlioptas process (AP), is that it discourages the 
growth of the larger clusters and encourages the smaller ones which inevitably delays the transition. 
Eventually, when it reaches near the critical point it is so unstable that occupation
of  one or two links triggers an explosion of growth. It leads to the emergence of a giant cluster with a bang
and hence it is called ``explosive percolation" (EP). Indeed, the corresponding $P$, in contrast to its 
classical counterpart, undergoes such an abrupt transition that it was at first mistaken as a discontinuity and 
suggested to exhibit the first order or discontinuous transition. Their results jolted the scientific community through a series 
of claims, unclaims and counter-claims  \cite{ref.Friedman, ref.ziff_1, ref.radicchi_1, ref.Costa_2,
ref.souza, ref.cho_1,  ref.ara, ref.da_Costa, ref.Grassberger, ref.Bastas}.
It is now well settled that the explosive percolation transition 
is actually continuous but with first order like finite-size effects 
 \cite{ ref.Grassberger, ref.Bastas, ref.Riordan, ref.bastas_review, ref.Choi}.

In general, scientists use theoretical model, just like architects use geometric model before building 
large expensive structure, because it provides useful insights into the real-world 
systems. The real systems that percolation represent is
complex as it often involves quantum and many particle interaction effects. 
However, modeling is only useful if we know how to relate
its various observable quantities to those of the real-world systems. To this end, 
Kasteleyn and Fortuin used the mapping of the percolation problem onto the $q$-state Potts
model in order to relate its observable to
the thermal quantities of the Potts model \cite{ref.Kasteleyn}. Owing to that mapping we know 
that $P$ is the order parameter,  mean cluster size 
$\langle s\rangle$  is the susceptibility etc. but not equivalent counterpart of entropy.
In thermal CPT, the entropy $S$ and the order parameter (OP) complement each other as $S$, that measures 
the degree of disorder, is maximum where OP is zero and OP, that measures the extent of order, is maximum
where $S$ is zero. A similar behaviour in percolation is also expected in order to elucidate whether it is also an order-disorder transition or not.
Universality is another aspect that we find common in the thermal CPT and in the random percolation.
In the case of EP, we are yet to find universality of any type or any kind. 
Another interesting aspect of thermal CPT is that its critical exponents $\alpha, \beta$
and $\gamma$ obey the  Rushbrooke inequality  $\alpha+2\beta+\gamma\geq 2$  
which reduces to equality under static scaling hypothesis \cite{ref.Stanley}. 
Whether it holds in explosive percolation or not, is also an interesting issue.

In this article, we investigate EP on the ER networks for product rule (PR) and sum rule (SR) and 
find their critical exponents numerically. First, we define susceptibility $\chi$ as the ratio 
of the successive jump $\Delta P$ of $P$ and the magnitude of successive intervals  $\Delta t$ instead of
using the mean cluster size $\langle s\rangle$ as susceptibility. Then we obtain
the critical exponents $\nu$ of the correlation length, $\gamma$ of $\chi$, and $\beta$ of $P$. Note that $\langle s\rangle$ exhibits the expected 
divergence only if the largest 
cluster size is excluded from it and even then it gives too large a value of $\gamma$. Realizing these drawbacks,
many researchers are already considering alternative definitions \cite{ref.radicchi_1, 
ref.ziff_3, ref.qian}. Second, we define entropy $H$ for EP and find that it is continuous across the whole spectrum
 of the
control parameter $t$ which clearly reveals that EP transition is indeed continuous in nature. We then 
define the specific heat as $C=q{{dH}\over{dq}}$ where $q=(1-t)$ and find that it diverges with positive critical exponent $\alpha$. 
The most intriguing and unexpected findings of this work is that PR and SR belong to the same universality class. Besides, we find that the elusive Rushbrooke inequality
holds in EP. Recently, using the the same definitions for entropy, specific heat and susceptibility we have shown that the Rushbrooke inequality holds in 
the random percolation too \cite{ref.hassan_didar}. Finding that RI also holds in EP on random network provides a clear testament of how robust our results are.

Percolation is all about clusters as every observable quantity of it is related, 
this way or another, to the clusters by virtue of definition.  
Initially, all the labeled nodes are considered isolated so that every node is a cluster of its own size. 
The process starts by picking two distinct links, say $e_{ij}$ and $e_{kl}$, randomly at each step. 
To apply the PR, we then calculate the products, $\Pi_{ij}=s_i\times s_j$ and $\Pi_{kl}= s_k \times s_l$,
of the size of the clusters that the two nodes on either side of each link contain.
The link with the smaller value of the products $\Pi_{ij}$ and $\Pi_{kl}$ is occupied.
On the other hand, if we find $\Pi_{ij}=\Pi_{kl}$ then  we occupy one of the two links at random 
with equal probability. In the case of SR, we take the sum $\Sigma_{ij}=s_i+s_j$ and $\Sigma_{kl}=s_k+s_l$ 
instead of the product and do the rest exactly in the same way as we did for PR. Each time we occupy a link, either 
the size of an existing cluster grows due to occupation of an inter-cluster link or the cluster size remains 
the same due to addition of an intra-cluster link. In either case, the growth of large clusters are always
disfavoured which is in sharp contrast to its  RP counterpart. Thus, the emergence of a giant cluster is considerably
slowed down but eventually when it happens, it happens abruptly but without discontinuity.

\begin{figure}

\centering

\subfloat[]
{
\includegraphics[height=4.0 cm, width=2.4 cm, clip=true,angle=-90]
{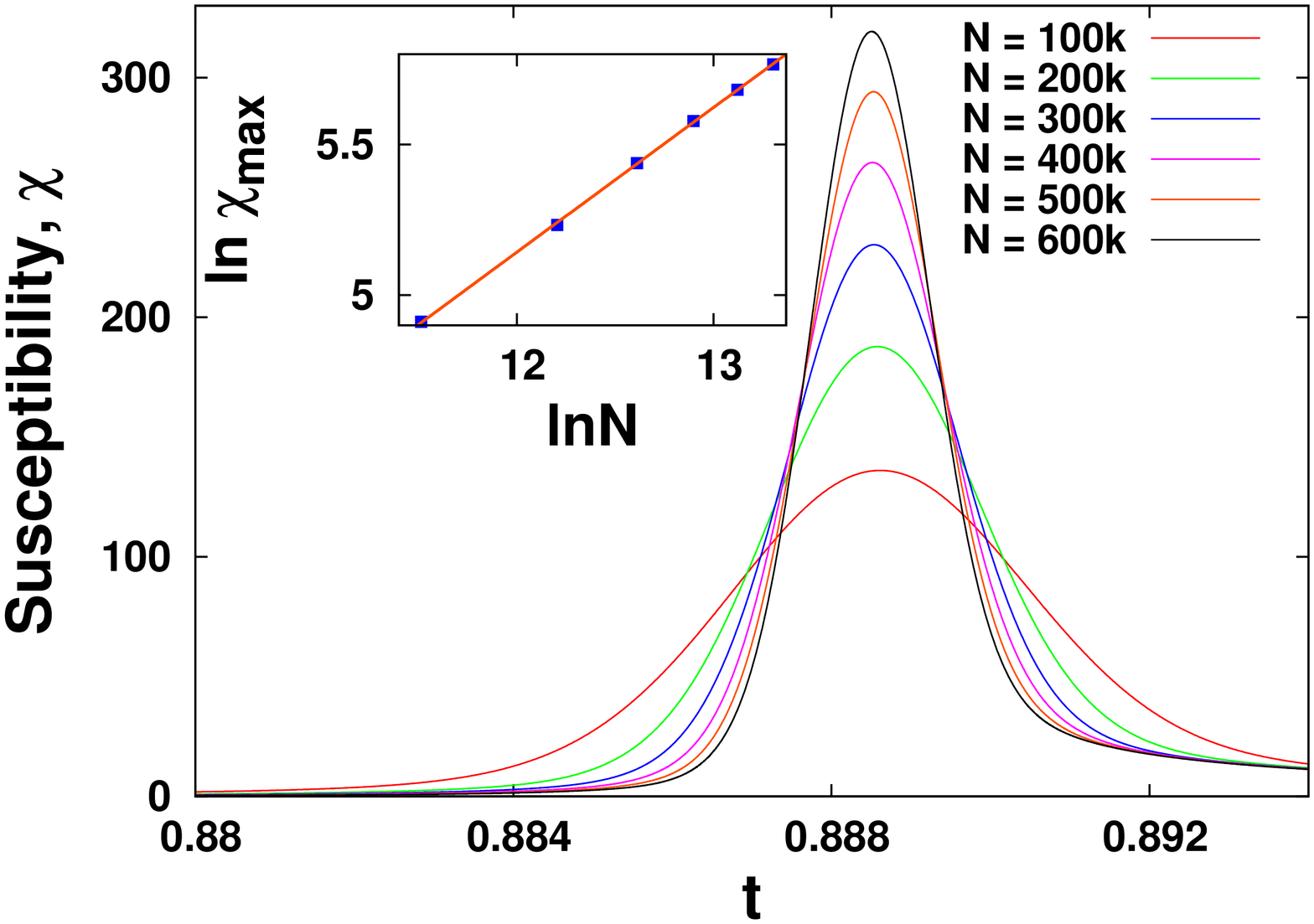}
\label{fig:1a}
}
\subfloat[]
{
\includegraphics[height=4.0 cm, width=2.4 cm, clip=true,angle=-90]
{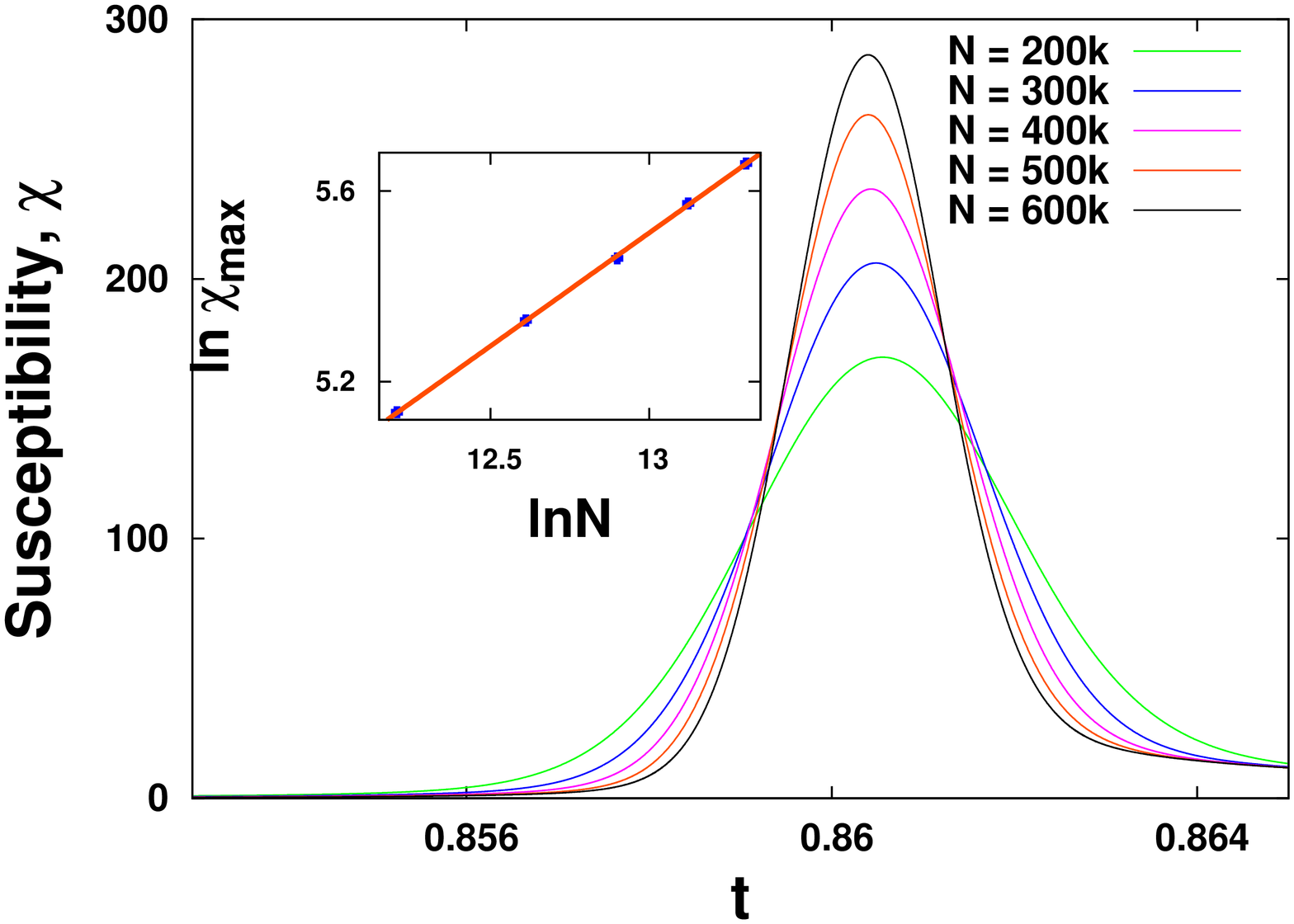}
\label{fig:1b}
}


\subfloat[]
{
\includegraphics[height=4.0 cm, width=2.4 cm, clip=true, angle=-90]
{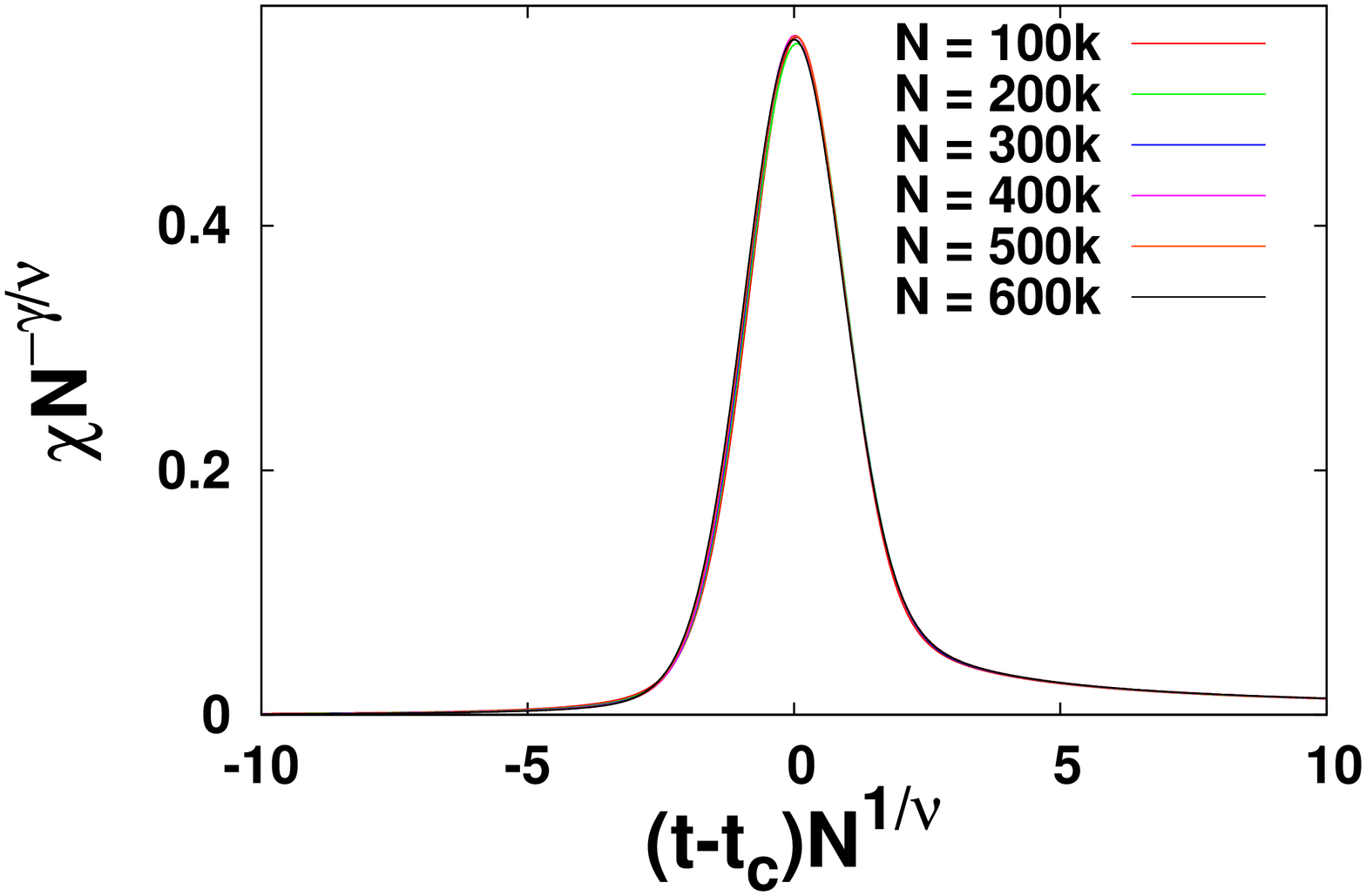}
\label{fig:1c}
}
\subfloat[]
{
\includegraphics[height=4.0 cm, width=2.4 cm, clip=true, angle=-90]
{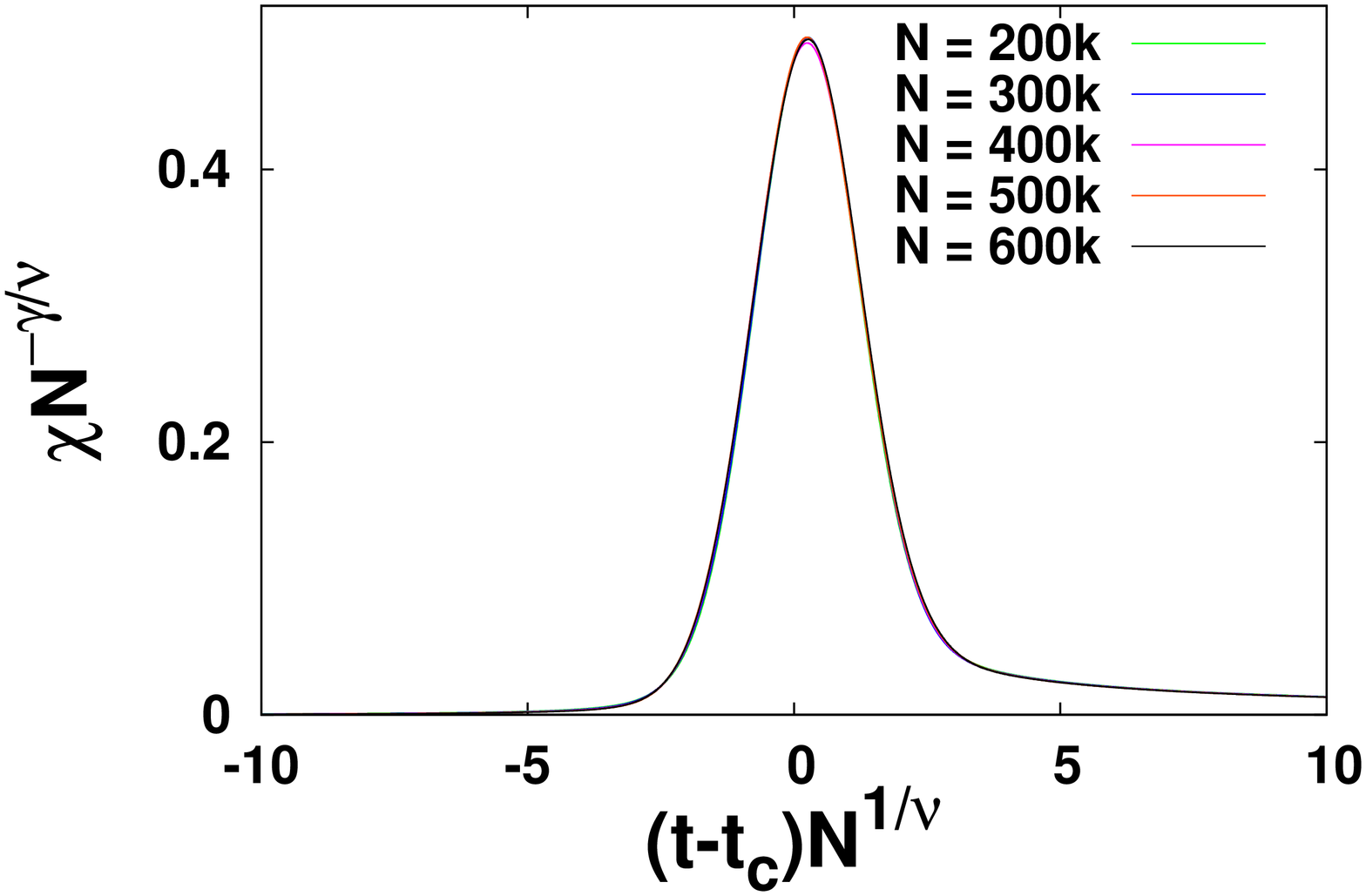}
\label{fig:1d}
}
\caption{
Susceptibility $\chi$ versus $t$ for PR and SR on the ER network is shown in (a) and (b) respectively. 
 In (c)-(d) we plot $\chi N^{-\gamma/\nu}$ vs $(t-t_c)N^{1/\nu}$ for PR and SR respectively and find that all the  
distinct plots for different network size collapse sharing the same critical exponents $\gamma/\nu=0.478$ and
$1/\nu=0.535$.
} 

\label{fig:1abcd}
\end{figure}

We first investigate not the $P$ itself but its successive jump $\Delta P$ within successive
interval $\Delta t=1/N$. The idea of successive jump size $\Delta P$ was first introduced by Manna \cite{ref.manna}. We use it to define susceptibility as 
\begin{equation} 
\chi(t)={{\Delta P}\over{\Delta t}},
\end{equation}
which essentially becomes the derivative of $P$ 
in the limit  $N\rightarrow \infty$. 
In Figs. (\ref{fig:1a}) and (\ref{fig:1b}) we show plots of $\chi$ versus $t$ for both 
PR and SR model. According to the finite-size scaling (FSS) hypothesis, the susceptibility  $\chi_{{\rm max}}$ 
at $t=t_c$ increases following a power-law $\chi_{{\rm max}}\sim N^{\gamma/\nu}$. In an attempt
to verify this we plot $\log(\chi_{{\rm max}})$ vs $\log(N)$,
see insets of Figs.  (\ref{fig:1a}) and (\ref{fig:1b}), and find straight lines
with slopes $\gamma/\nu=0.480(3)$ for PR and $\gamma/\nu=0.475(4)$ for SR. 
Following the procedures in Ref. \cite{ref.Hassan_Rahman_1} we also get a rough 
estimate of the exponent $1/\nu = 0.535(5)$ for PR and $1/\nu =  0.537(1)$ for SR. 
The FSS theory further suggests that if we now plot $\chi N^{-\gamma/\nu}$ vs 
$(t_c-t)N^{1/\nu}$, all the
distinct plots of Figs. (\ref{fig:1a}) and (\ref{fig:1b}) should collapse into their respective universal
curves. Indeed, by tuning ${\gamma/\nu}$ and $1/\nu$ we find excellent data collapse, see Figs. (\ref{fig:1c}) and (\ref{fig:1d}), 
if we use $\gamma/\nu=0.478$ and $1/\nu=0.535$ for PR
and SR respectively. Note that $t_c$ value also affects the data collapse and hence tuning the initial estimates for $t_c$ we get the best data-collapse if we use 
$t_c=0.88850$ for PR and $t_c= 0.86018$ for SR.
The quality of data collapse itself provides a clear testament to the extent of accuracy of these values. What is most noteworthy, however, is that both PR and SR share the same value for
the exponents $\gamma$ and $\nu$. Using now the relation $N\sim (t-t_c)^{-\nu}$ in $\chi\sim N^{\gamma/\nu}$ 
we find that 
\begin{equation}
\chi\sim (t-t_c)^{-\gamma},
\end{equation}
where $\gamma=0.893$ for both PR and SR rules within the acceptable limit of error. 
We find that the susceptibility now diverges even without the exclusion of the largest cluster and that too
with the same $\gamma$.

\begin{figure}

\centering

\subfloat[]
{
\includegraphics[height=4.0 cm, width=2.4 cm, clip=true,angle=-90]
{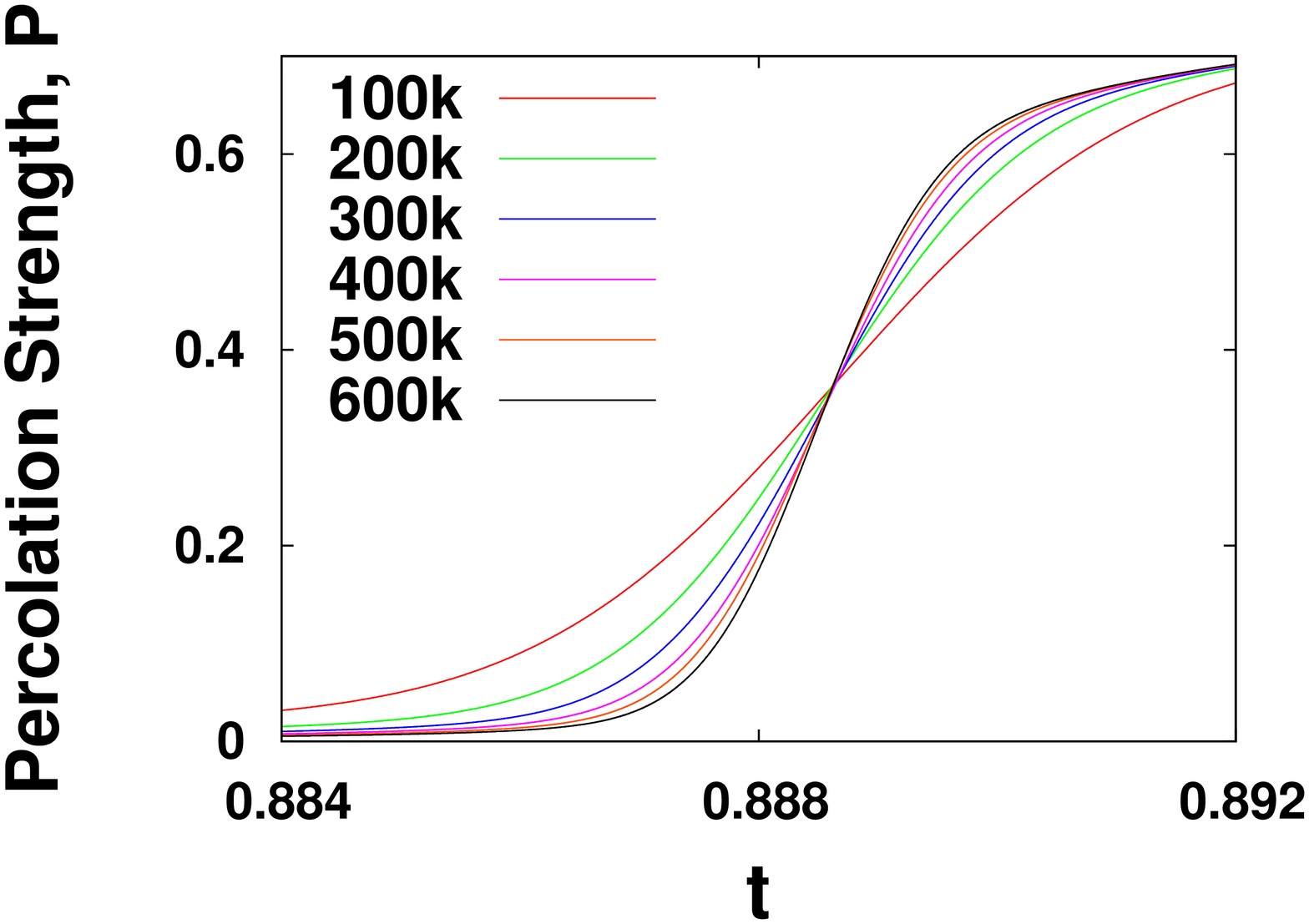}
\label{fig:2a}
}
\subfloat[]
{
\includegraphics[height=4.0 cm, width=2.4 cm, clip=true,angle=-90]
{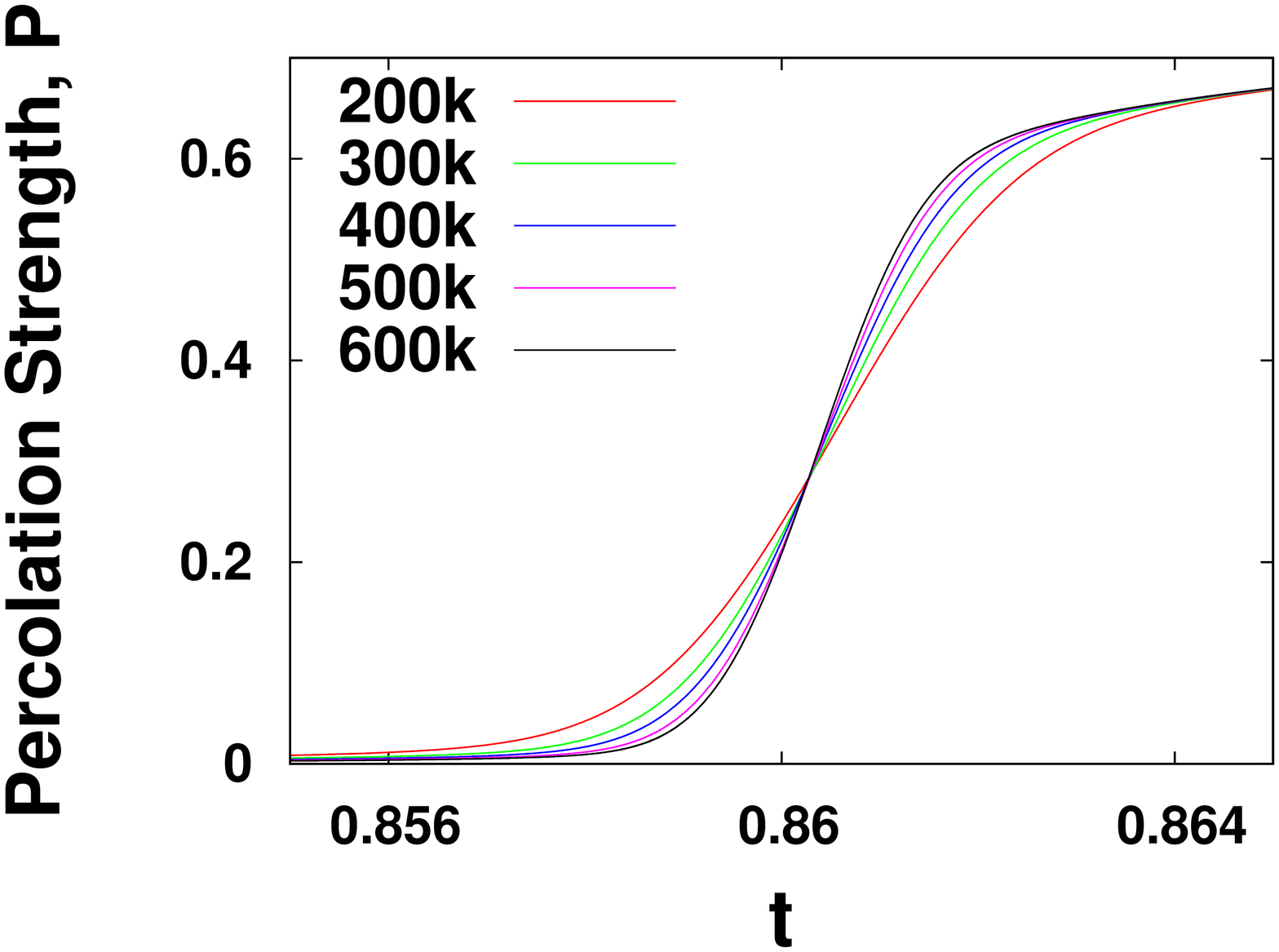}
\label{fig:2b}
}

\subfloat[]
{
\includegraphics[height=4.0 cm, width=2.4 cm, clip=true, angle=-90]
{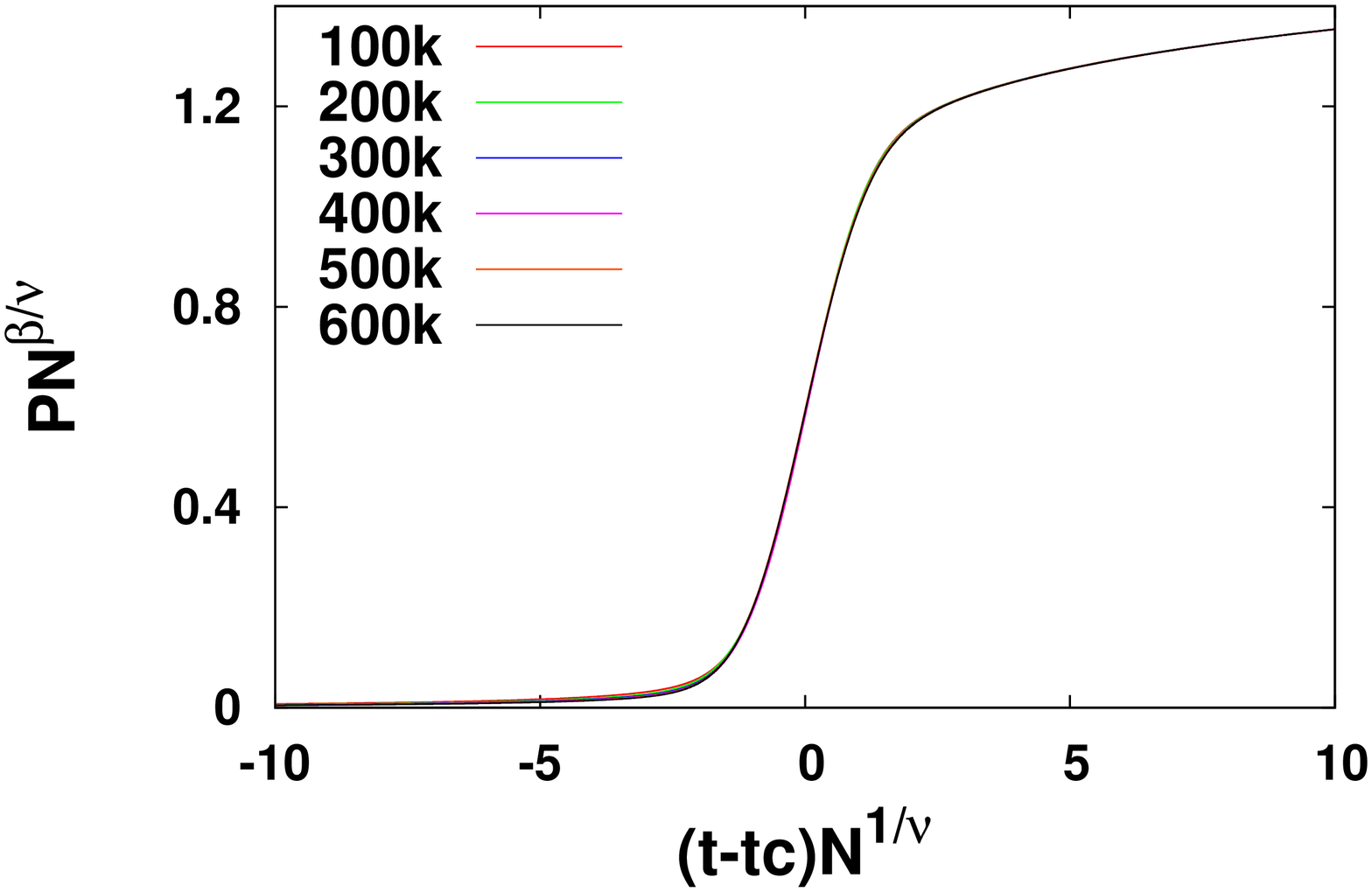}
\label{fig:2c}
}
\subfloat[]
{
\includegraphics[height=4.0 cm, width=2.4 cm, clip=true, angle=-90]
{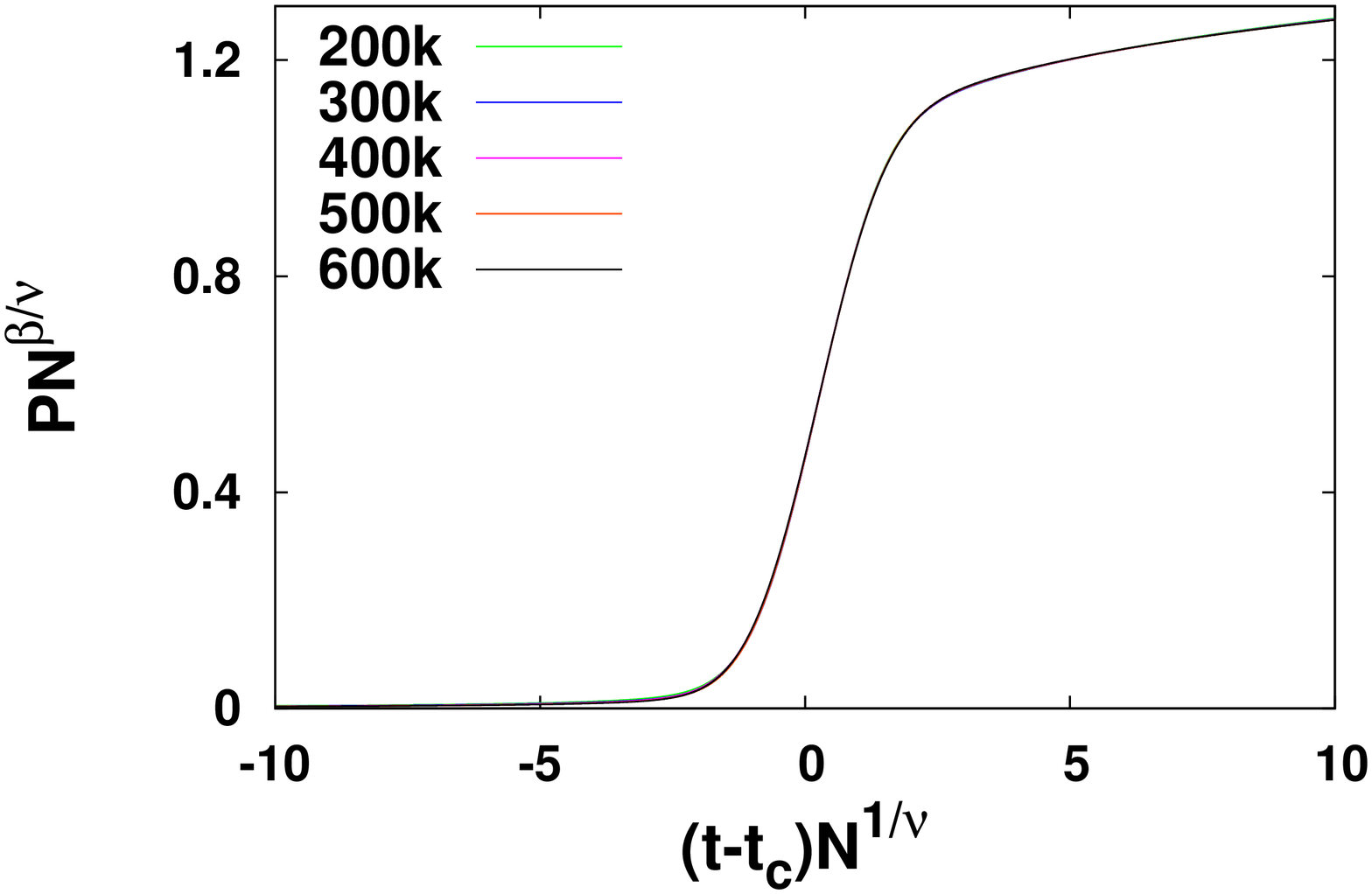}
\label{fig:2d}
}
\caption{Plots of order parameter $P$ versus $t$ for EP of PR in (a) and SR in (b). We plot $PN^{\beta/\nu}$ versus $(t-t_c)N^{1/\nu}$ and
find that all the distinct plots of (a) and (b) collapse superbly in (c) for PR and (d) for SR.
} 

\label{fig:2abcd}

\end{figure}

Now, we consider the order parameter $P$ itself and
plot it as a function of $t$ in Figs. (\ref{fig:2a}) and (\ref{fig:2b}) for 
PR and SR respectively. We follow the same standard procedure as in 
Ref. \cite{ref.Hassan_Rahman_1, ref.Hassan_Rahman_explosive} and find  $\beta/\nu=0.045$  for both the variants.
It is well-known that $P(t,N)$ exhibits finite-size scaling.
One way of testing it is 
to plot  $PN^{\beta/\nu}$ vs $(t-t_c)N^{1/\nu}$ and check if all the distinct 
curves of $P$ vs $t$ curves collapse or not.
Indeed, Figs. (\ref{fig:2c}) and (\ref{fig:2d}) suggest that they all collapse superbly  with 
$\beta/\nu=0.045$ and $1/\nu=0.535$ values regardless of whether it is PR or SR. 
Substituting the relation $N\sim (t-t_c)^{-\nu}$ in $P\sim N^{-\beta/\nu}$ we get
\begin{equation}
P(t)\sim (t-t_c)^\beta.
\end{equation}
This is exactly how the order parameter behaves near critical point in the thermal CPT as well. 
We once again find that both PR and SR rules share the same
exponent $\beta =0.084$ within the acceptable limits of error. Such unusually 
low value of $\beta$ compared to that of the RP on ER where $\beta=1$ is the hallmark of EP transition
\cite{ref.mori}. Note also that Grassberger {\it et al.}  obtained $1/\nu=0.5$
and $\beta=0.0861(5)$ for PR on ER \cite{ref.Grassberger}. Our values are quite close to their 
values; however little differences are there which marks a significant 
improvement in the quality of data-collapse.

\begin{figure}

\centering

\subfloat[]
{
\includegraphics[height=4.0 cm, width=2.4 cm, clip=true,angle=-90]
{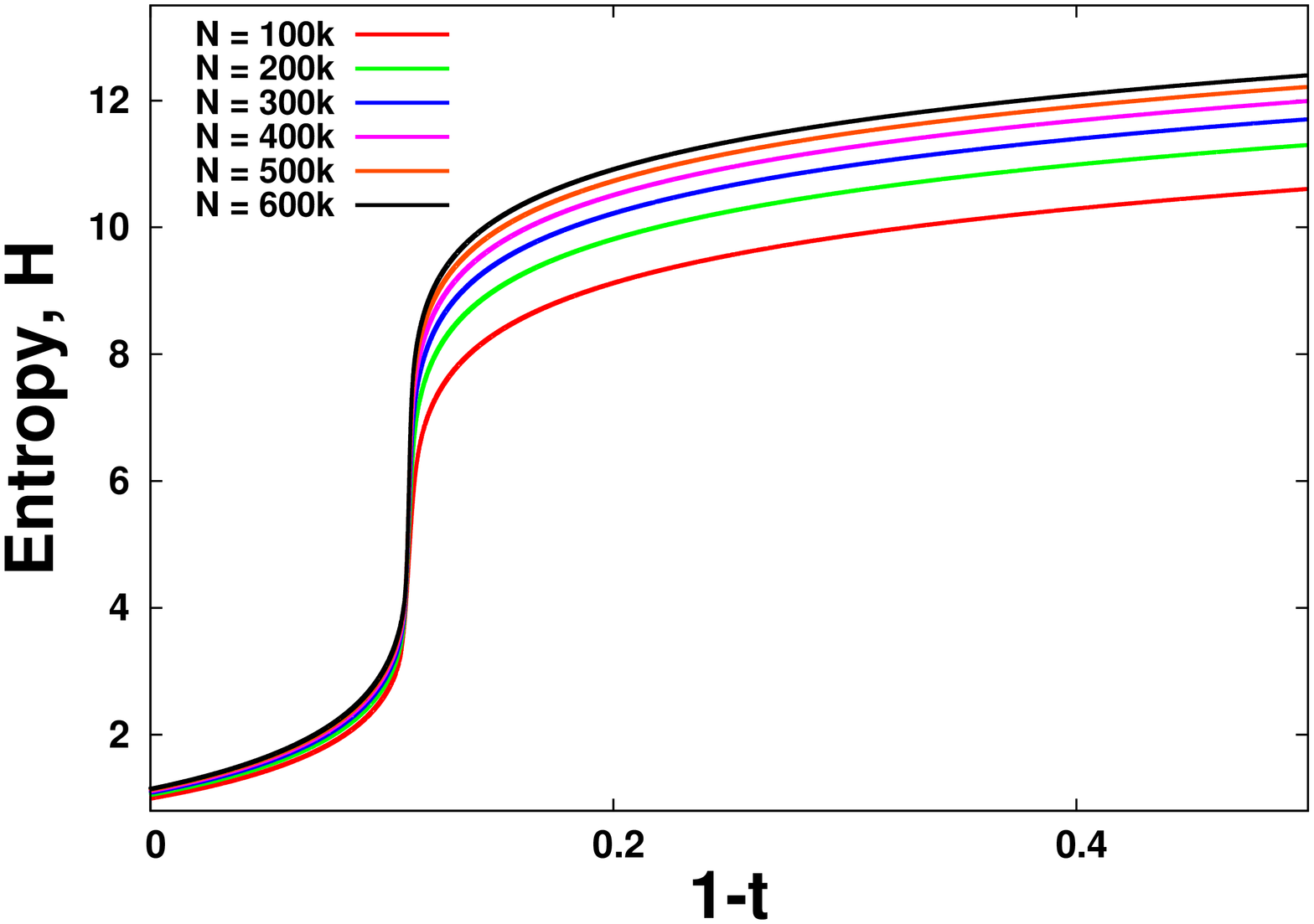}
\label{fig:3a}
}
\subfloat[]
{
\includegraphics[height=4.0 cm, width=2.4 cm, clip=true,angle=-90]
{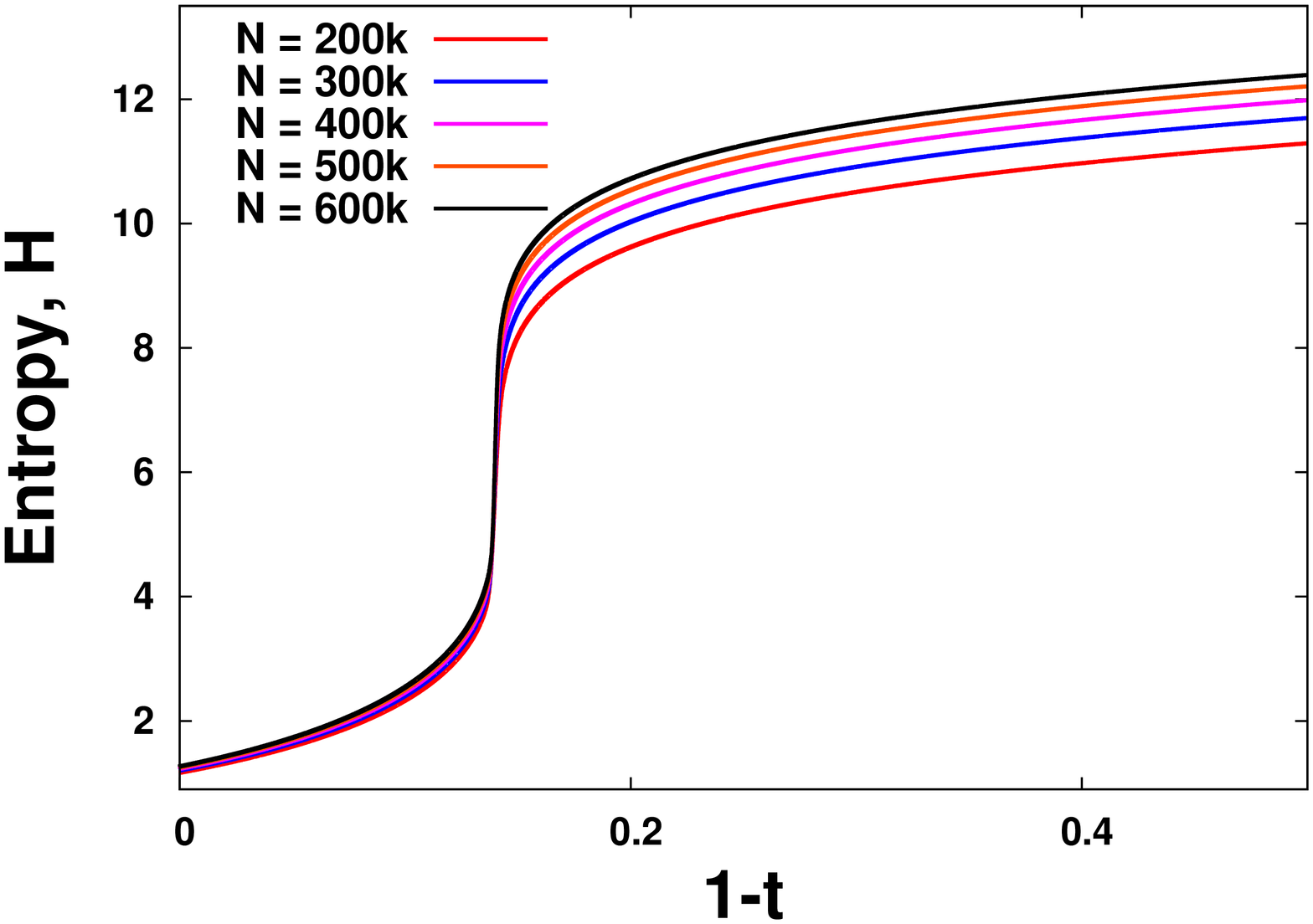}
\label{fig:3b}
}

\caption{Entropy $H$ versus $1-t$ for explosive percolation on the ER network following (a) PR and (b)
 SR of the Achlioptas process. 
} 

\label{fig:3abcd}
\end{figure}

Phase transitions always entail a change in entropy and hence no model for phase transition is complete 
without a proper definition for it. To this end, we find that the most suitable choice for entropy in percolation
is the Shannon entropy which is defined as
\begin{equation}
\label{eq:shannon_entropy}
H(t)=-K\sum_i^m \mu_i\log \mu_i,
\end{equation} 
where we choose $K=1$ since it merely amounts to a choice of a unit of measure of entropy \cite{ref.shannon}. 
Although there is no explicit restriction per se on the choice of $\mu_i$
there exist some implicit restrictions. 
The text-book definitions of thermal entropy $S$ and the specific heat $c$
suggest that the $S$ vs $T$ plot must always have a sigmoidal shape with positive slope \cite{ref.Stanley}.
Recently, Vieira {\it et al.}  used the probability $w_s$, that  
a node picked at random belongs to a cluster exactly of size $s$, in Eq. (\ref{eq:shannon_entropy}) 
to measure Shannon entropy for explosive percolation and found
that the entropy increases from zero at $t=0$ to reach its maximum value at $t_c$ followed by sharp decrease above $t_c$ \cite{ref.Vieira}. We also
know that the order parameter is also zero at $t=0$. It means that 
the system is ordered and disordered at the same time which is not possible. Besides, the
bell-shaped like entropy curve also violates the second law of thermodynamics. 
The problem lies in the fact that the sum in Eq. (\ref{eq:shannon_entropy}) is over each individual cluster
not over a class of cluster of size $s$ and hence one cannot use $w_s$ to measure entropy. Note that the Shannon entropy measures how much information 
is contained in each 
cluster like in each message in the information theory. To find the appropriate probability $\mu_i$ for
Eq. (\ref{eq:shannon_entropy}), we assume that for a given
$t$ there are $m$ distinct and disjoint labeled clusters $i=1,2,...,m$ of size $s_1,s_2,....,s_m$ 
respectively. We then propose  a labeled cluster picking probability (CPP) $\mu_i$, that a 
node picked at random belongs to cluster $i$, and assume that it depends on the size $s_i$ of the cluster $i$ itself,
so that $\mu_i=s_i/\sum_i s_i$ where $\sum_i s_i=N$.

Incorporating $\mu_i=s_i/N$ in Eq. (\ref{eq:shannon_entropy}) 
we obtain entropy for explosive percolation. To visualize we plot it in Figs. (\ref{fig:3a}) 
and (\ref{fig:3b}) as a function of $q=1-t$ for PR and SR respectively. We observe that 
the maximum entropy occurs at $q=1$ where $\mu_i=1/N$ $\ \forall \ i$ which means that every node has 
the same probability to be picked if we hit one at random.
This is exactly like the state of the isolated ideal gas since here too all accessible microstates 
are equally probable. The $q=1$ state is thus the most confused or disordered state.
Now as we lower the $q$ value, we see that entropy decreases slowly 
but as we approach towards $q_c=1-t_c$ we observe a dramatic decrease in entropy. 
This is because  as we approach $q_c$ from higher $q$ value 
we find that many moderately large sized clusters get accumulated as the AP discourages growth of large clusters 
and encourages the smaller ones. Eventually the crowding of the moderately large clusters reaches a critical 
state at $q=q_c$ where addition of a few links 
triggers the growth of the largest cluster in an explosive fashion. We find that at $q=0$ the entropy $H$ is minimally low but the order parameter $P$ is maximally high and 
hence it is clearly the ordered phase.
We thus see that at $q=1$ the entropy is maximally high but the order parameter $P=0$ and 
hence it correspond to the disordered phase.  On the other hand at $q=0$, the order parameter is maximally
high and entropy is minimally low. 
The term percolation therefore refers to the transition 
from ordered phase characterized by vanishingly small entropy at $q<q_c$ to disordered phase characterized
by $P=0$ at $q>q_c$ as one tunes the control parameter $q$. We thus find that in percolation
too, like in the thermal CPT, entropy $H$ and order parameter $P$ compliments each other.

\begin{figure}

\centering

\subfloat[]
{
\includegraphics[height=4.0 cm, width=2.4 cm, clip=true,angle=-90]
{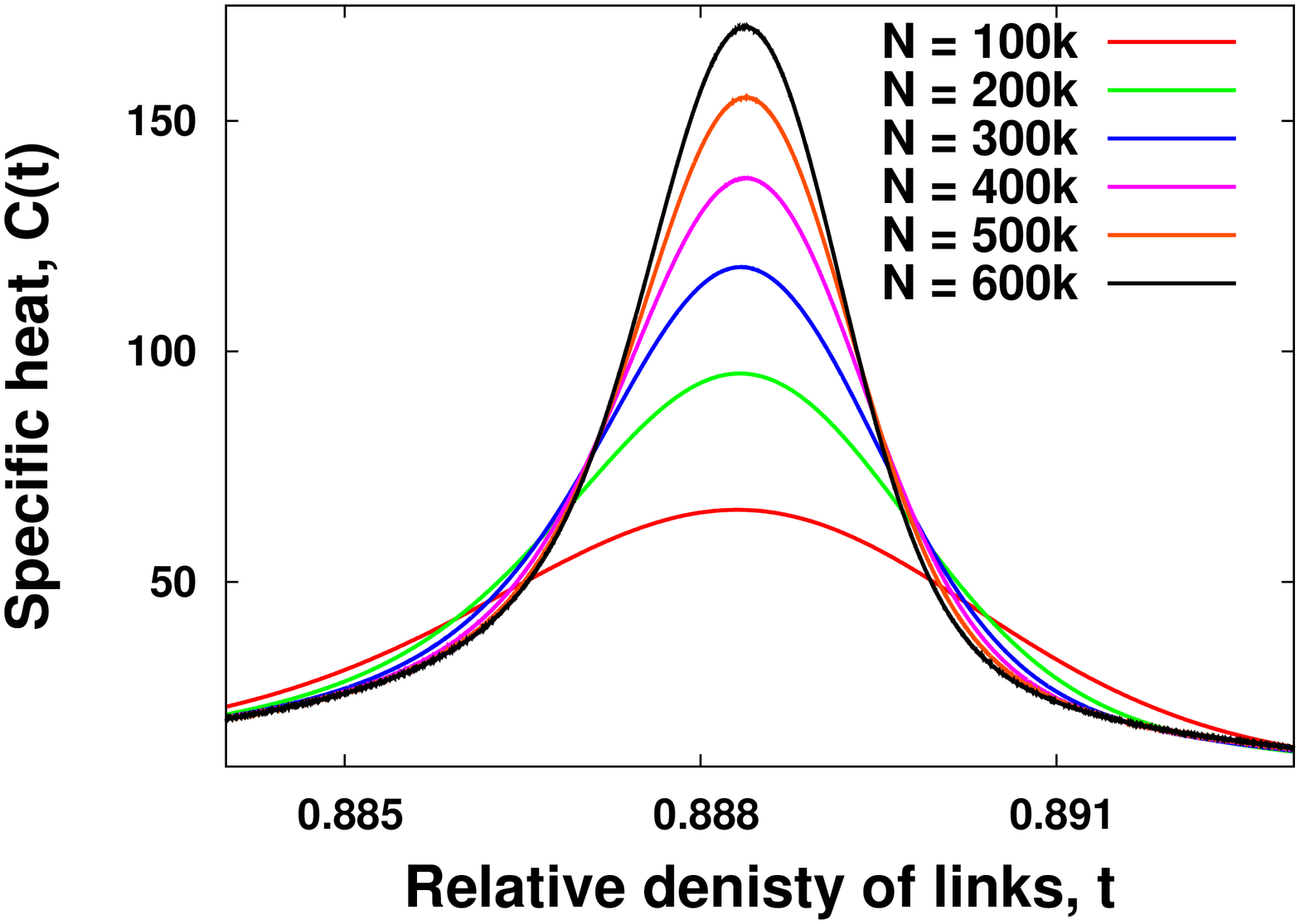}
\label{fig:4a}
}
\subfloat[]
{
\includegraphics[height=4.0 cm, width=2.4 cm, clip=true,angle=-90]
{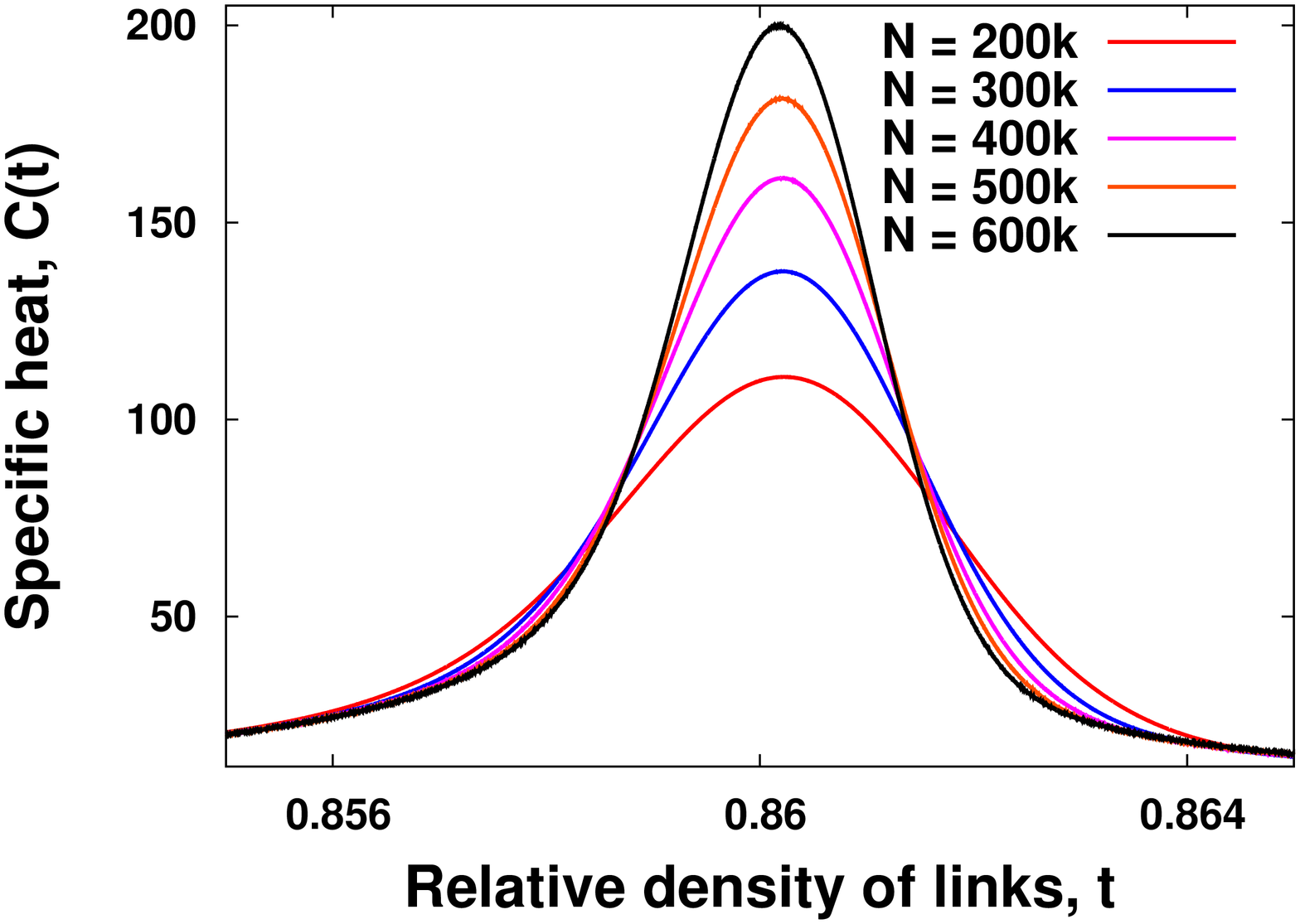}
\label{fig:4b}
}

\subfloat[]
{
\includegraphics[height=4.0 cm, width=2.4 cm, clip=true, angle=-90]
{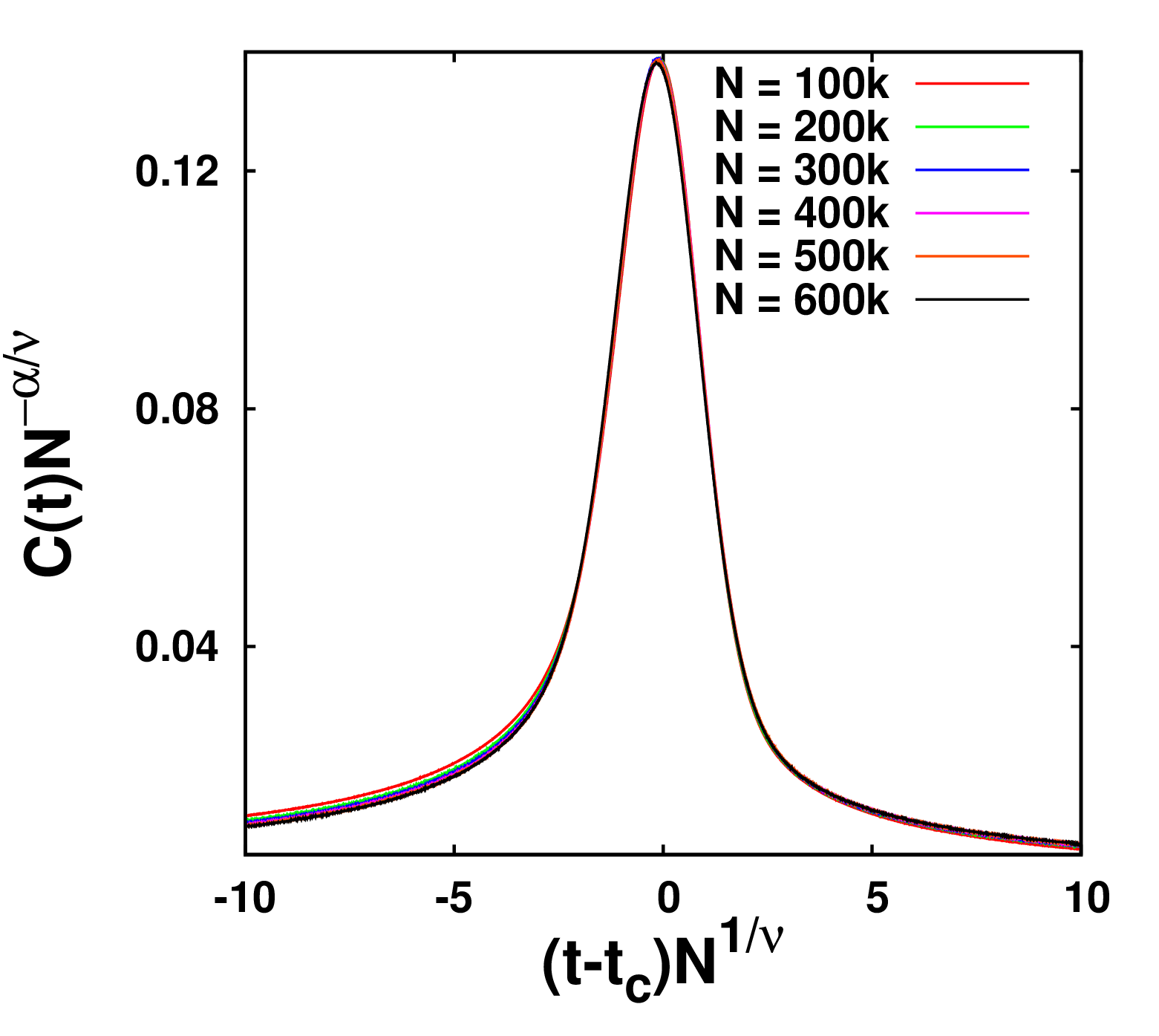}
\label{fig:4c}
}
\subfloat[]
{
\includegraphics[height=4.0 cm, width=2.4 cm, clip=true, angle=-90]
{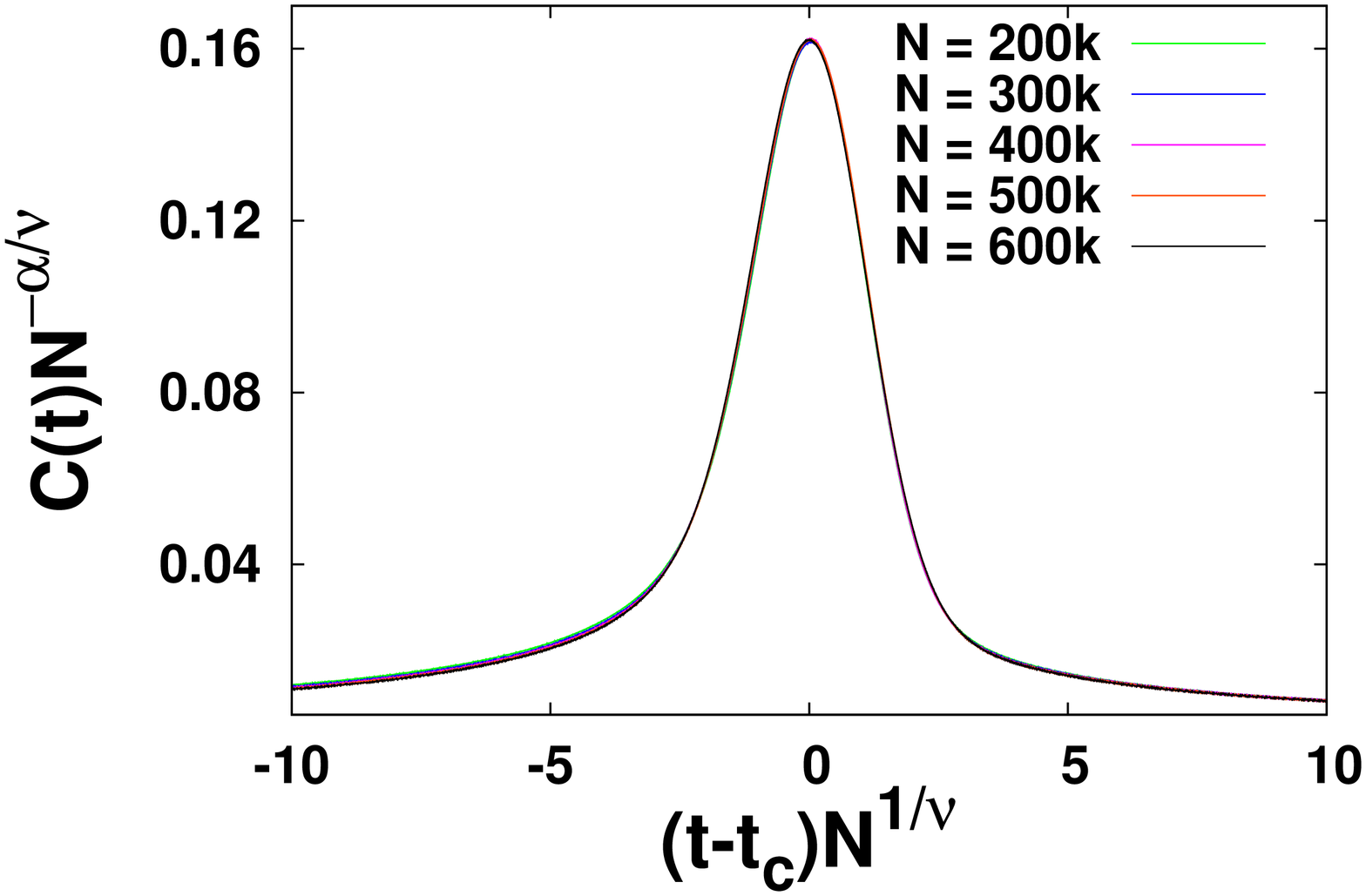}
\label{fig:4d}
}
\caption{Plots of specific heat $C$ for PR and SR are shown in (a) and (b) respectively as a function of $t$. 
In (c) and (d) we plot t $CN^{-\alpha/\nu}$ vs
$(t-t_c)L^{1/\nu}$ and find that all the distinct plots of (a) and (b) collapse superbly sharing the same 
$\alpha$ and $\nu$ values.
} 

\label{fig:4abcd}
\end{figure}



Once we know the entropy, we can find specific heat as we can define it as
\begin{equation}
C(t)=(1-t){{dH}\over{d(1-t)}}
\end{equation}
 in analogy with the definition of its thermal counterpart. 
Taking differentiation of $H$ from first principles and multiplying that value with the corresponding value of $(1-t)$,
we can immediately obtain $C(t)$. We then plot $C(t)$ in Figs. (\ref{fig:4a}) and (\ref{fig:4b}) as a function
of $t$ for PR and SR respectively. 
To compute the corresponding critical exponent $\alpha$ once again we use the FSS hypothesis
 and find $\alpha/\nu=0.535$ for both PR and SR. Finally, we plot $CL^{-\alpha/\nu}$ vs $(t-t_c)L^{1/\nu}$ 
and obtain a perfect data-collapse with $\alpha/\nu=0.535$ and $1/\nu=0.535$ for
both PR and SR as shown in Figs. (\ref{fig:4c}) and (\ref{fig:4d}).  
We then use the relation  $L\sim (t-t_c)^{-\nu}$ in $C(t)\sim L^{\alpha/\nu}$ and immediately 
find that the specific heat diverges like
\begin{equation}
C(t)\sim (t-t_c)^{-\alpha},
\end{equation} 
where $\alpha=1$ for both PR and SR. The quality of data-collapse is a clear testament of  
the accuracy of $\alpha$ value.

Classifications of any system into universality classes
is always an interesting proposition.
To this end, finding that PR and SR of explosive percolation belongs to the
same universality class is a significant development
 especially when we know that the most expected site-bond universality breaks down even in the lattice.
To check whether the Rushbrooke inequality holds in EP or not, we
 substitute our values of $\alpha=1$, $\gamma=0.893$ and $\beta=0.084$ in the Rushbrooke relation and
find $\alpha+2\beta+\gamma=2.061$. 
Thus, we find that the Rushbrooke inequality not only holds but also its
value is close to equality, within the acceptable range of errors. 
Recently, we applied the same approach to the square and weighted planar stochastic (WPS) lattices
where we found that RI holds in RP for both the lattices albeit they belong to different universality classes \cite{ref.hassan_didar}. 
Moreover, in both the cases, we find that RI holds almost as an equality like in the thermal CPT. Thus, finding that 
RI holds for three different 
universality classes that include a class as exotic as EP provides sufficient confidence in our results. 
It implies that explosive percolation is indeed a paradigmatic model for continuous phase transition with
some unusual finite-size behaviours since we find hysteresis loops in its forward and reverse processes, 
doublehump in the distributions of the order parameter $P$, which, however, disappears in the thermodynamic limit \cite{ ref.Grassberger, ref.Bastas, ref.Riordan, ref.bastas_review, ref.Choi}. 
Besides, we also know that the time difference $\Delta=t_2-t_2$ between the last step $t_2$ for which the 
largest cluster $C <N^{1/2}$ and the first step $t_1$ for which $C>0.5N$ is not extensive while it is extensive for RP on ER . 
For all these reasons explosive percolation is indeed a non-trivial paradigmatic model for CPT.

To summarize, we have used our recently defined entropy, specific heat and re-defined susceptibility in 
explosive percolation on random network. Until now 
we could only quantify the extent of order in percolation by measuring the order parameter $P$. 
 Thanks to the definition of entropy, we can now quantify 
the other phase too.  It is so high in the phase where $P=0$ that we can regard it as disordered phase.
It implies that the high-$q$ phase is more disordered, i.e., has a higher symmetry than the low-$q$ phase thus revealing 
that percolation is an order-disorder transition like
ferromagnetic transition. We have also shown that the specific heat and susceptibility diverge at the critical point 
without having to exclude the largest cluster which is in sharp contrast to the mean cluster size 
which also diverges at the critical point but only if we exclude the
largest cluster from it. We obtained the critical exponents $\alpha$, $\beta$, $\gamma$  and $\nu$ numerically 
and found that their values  for PR and SR are the same, revealing that they belong to the
same universality class. Such PR-SR universality is highly intriguing and unexpected especially
against the background of the breakdown of the usual site-bond universality even in the lattice.
We have also shown that the value of the critical exponents $\alpha$, $\beta$, $\gamma$ 
obey the Rushbrooke inequality. Our work confirms that the explosive percolation model is
a truly paradigmatic model for continuous phase transition since we now know that entropy, order parameter, specific heat, susceptibility and their critical exponents
behave exactly in the same way as  in the thermal CPT.

\end{document}